\documentclass[a4paper]{jpconf}

\usepackage{amssymb}
\usepackage{amsmath}

\usepackage{latexsym}
\usepackage{graphicx}

\usepackage{cite}

\newcommand{\be}{\begin{equation}}
\newcommand{\ee}{\end{equation}}

\newcommand{\dlt}{\delta}

\newcommand{\bbr}{{\bf r}}

\newcommand{\bbe}{{\bf e}}
\newcommand{\bB}{{\bf B}}
\newcommand{\bS}{{\bf S}}
\newcommand{\bn}{{\bf n}}
\newcommand{\bt}{\beta}
\newcommand{\vp}{\varphi}

\newcommand{\al}{\alpha}

\newcommand{\gm}{\gamma}
\newcommand{\om}{\omega}

\newcommand{\lbd}{\lambda}

\newcommand{\rgl}{\rangle}
\newcommand{\lgl}{\langle}

\begin{document}

\title{Spin superradiance by magnetic nanomolecules and nanoclusters}

\author{V I Yukalov$^1$, V K Henner$^{2,3}$ and E P Yukalova$^4$}

\address{$^1$Bogolubov Laboratory of Theoretical Physics,
Joint Institute for Nuclear Research, \\ Dubna 141980, Russia \\
$^2$Department of Physics, Perm State University, Perm 614190, Russia \\
$^3$Department of Physics, University of Louisville, Louisville,
Kentucky 40292, USA \\
$^4$Laboratory of Information Technologies,
Joint Institute for Nuclear Research, \\ Dubna 141980, Russia
}

\ead{yukalov@theor.jinr.ru}

\begin{abstract}
Spin dynamics of assemblies of magnetic nanomolecules and nanoclusters can
be made coherent by inserting the sample into a coil of a resonant electric
circuit. Coherence is organized through the arising feedback magnetic field
of the coil. The coupling of a magnetic sample with a resonant circuit induces
fast spin relaxation and coherent spin radiation, that is, superradiance.
We consider spin dynamics described by a realistic Hamiltonian, typical of
magnetic nanomolecules and nanoclusters. The role of magnetic anisotropy is
studied. A special attention is paid to geometric effects related to the
mutual orientation of the magnetic sample and resonator coil.

\end{abstract}

\section{Introduction}

There exists a large class of magnetic nanomolecules and magnetic
nanoclusters that can be considered as nanoparticles possessing high
total spins (see review articles
\cite{Barbara_1,Wernsdorfer_2,Ferre_3,Yukalov_4,Yukalov_5,Bedanta_6,
Berry_7,Beveridge_8,Hoang_9}). Below blocking temperature, the spin of such
magnetic nanoparticles is frozen. For instance, the typical blocking
temperature of magnetic nanomolecules is of order $1 - 10$ K. The blocking
temperature for nanoclustres is $10 -100$ K.

Magnetic properties of nanomolecules and nanoclusters are similar to each
other. There are two main features distinguishing them. Magnetic molecules
of the same chemical composition are identical and they can form crystals with
almost ideal periodic lattice. While magnetic nanoclusters, even being made
of the same element, say Fe, Ni, or Co, differ by their sizes, and they do
not form periodic structures. Otherwise, the spin Hamiltonian for an ensemble
of magnetic nanoparticles is of the same form for nanomolecules as well as
for nanoclusters.

In the usual case, spin relaxation is due to spin-phonon interactions and,
below the blocking temperature, is very slow. Thus for nanomolecules, the
spin-phonon relaxation time is $T_1 \sim (10^5 - 10^7)$ s. But the spin
relaxation time can be drastically shortened, if the magnetic sample is inserted
into a coil of a resonant electric circuit. This is termed the Purcell
effect \cite{Purcell_10}. In that case, the relaxation is caused by the
resonator feedback field collectivizing moving spins and forcing them to move
coherently. Coherent spin dynamics have been studied in several publications,
e.g., \cite{Kiselev_11,Belozerova_12,Belozerova_13,Yukalov_14,Yukalov_15,
Yukalov_16,Yukalov_17,Yukalov_18,Belozerova_19,Yukalov_20,Davis_39,Yukalov_40,
Yukalov_21,Yukalov_22,Henner_23}.

Coherently moving spins produce coherent radiation, which, when it is
self-organized, is called superradiance. It is worth stressing that spin
superradiance is rather different from atomic superradiance. The latter is
caused by the Dicke effect \cite{Dicke_24}, while the cavity Purcell effect is
secondary \cite{Walther_25,Manassah_26,Manassah_27}. Contrary to this, spin
superradiance is completely due to Purcell effect, with the Dicke effect playing
no role \cite{Yukalov_28}. The Purcell effect also enhances the signals of
nuclear magnetic resonance \cite{Chen_29,Krishnan_30} and of spin echo
\cite{Shavishvili_31,Fokina_32,Nazarova_33}.

Here we consider the peculiarities of spin superradiance by magnetic
nanomolecules and nanoclusters having strong magnetic anisotropy. We shall pay
attention to the role of geometric effects related to the finiteness of the
considered samples. Finite systems, as is known \cite{Jin_34,Birman_35}, can
exhibit properties different from those of bulk systems. In the present case,
we are interested in the geometric effects due to the mutual orientation of a
finite magnetic sample and the resonator coil.

\section{Spin Hamiltonian}

An ensemble of magnetic nanomolecules or nanoclusters is described by the
Hamiltonian
\be
\label{1}
\hat H = \sum_i \hat H_i + \frac{1}{2} \sum_{i\neq j} \hat H_{ij} \;   ,
\ee
consisting of single-spin terms $\hat{H}_i$ and spin-interaction terms
$\hat{H}_{ij}$, with the index $i = 1,2,\ldots, N$ enumerating nanoparticles.
The single-spin Hamiltonian
\be
\label{2}
\hat H_i = - \mu_0 \bB\cdot\bS - D ( S_i^z)^2 + D_2 (S_i^x)^2 +
D_4 \left [ (S_i^x)^2 (S_i^y)^2 + (S_i^y)^2 (S_i^z)^2 +
(S_i^z)^2 (S_i^x)^2 \right ]
\ee
is a sum of the Zeeman energy and single-site magnetic anisotropy terms.
The total magnetic field, acting on each spin,
\be
\label{3}
 \bB = B_0 \bbe_z + H \bbe_x  ,
\ee
includes an external field $B_0$ and a resonator feedback field $H$. Spins
interact with each other through dipolar forces characterized by the
Hamiltonian
\be
\label{4}
 \hat H_{ij} = \sum_{\al\bt} D_{ij}^{\al\bt} S_i^\al S_j^\bt \;  ,
\ee
with the dipolar tensor
$$
 D_{ij}^{\al\bt} = \frac{\mu_0^2}{r_{ij}^3} \; \left ( \dlt_{\al\bt} -
3 n_{ij}^\al n_{ij}^\bt \right ) \;  ,
$$
where
$$
r_{ij} \equiv | \bbr_{ij} | \; , \qquad
\bn_{ij} \equiv \frac{\bbr_{ij}}{r_{ij}} \; ,
\qquad \bbr_{ij} \equiv \bbr_i - \bbr_j \;   .
$$

The resonator feedback field is given by the Kirchhoff equation
\be
\label{5}
\frac{dH}{dt} + 2\gm H + \om^2 \int_0^t H(t') \; dt' =
 - 4\pi\eta \; \frac{dm_x}{dt} \;  ,
\ee
in which $\gamma$ is resonator damping, $\omega$ is resonator natural
frequency, $\eta$ is filling factor, and
\be
\label{6}
m_x \equiv \frac{\mu_0}{V} \sum_{j=1}^N \lgl S_j^x \rgl
\ee
is the transverse magnetization density of the sample having volume $V$.

In addition to the resonator natural frequency $\omega$, there are the
following characteristic frequencies. The Zeeman frequency
\be
\label{7}
\om_0 \equiv - \; \frac{\mu_0}{\hbar} \; B_0 = \frac{2}{\hbar} \; \mu_B B_0
\ee
and the anisotropy frequencies
\be
\label{8}
 \om_D \equiv ( 2S - 1) \; \frac{D}{\hbar} \; , \qquad
\om_2 \equiv ( 2S - 1) \; \frac{D_2}{\hbar} \; , \qquad
\om_4 \equiv ( 2S - 1) \; \frac{D_4}{\hbar} \; S^2 \;  .
\ee
The resonator natural frequency has to be close to the Zeeman frequency,
in order to satisfy the resonance condition
\be
\label{9}
 \left | \frac{\om-\om_0}{\om} \right | \ll 1 \;  .
\ee
And the Zeeman frequency has to be larger than the anisotropy frequencies
that freeze spin motion,
\be
\label{10}
 \left | \frac{\om_D}{\om_0} \right | \ll 1 \; , \qquad
\left | \frac{\om_2}{\om_0} \right | \ll 1 \; , \qquad
\left | \frac{\om_4}{\om_0} \right | \ll 1 \; .
\ee
Among the anisotropy frequencies, the most important are $\omega_D$ and
$\omega_2$ that are close to each other. The frequency $\omega_4$, up to
spins $S \sim 10^3$, is much smaller than $\omega_D$.

We analyze the Heisenberg equations of motion for spins in two ways, by
employing the scale separation approach \cite{Yukalov_4,Yukalov_5} and by
directly solving the spin evolution equations in semiclassical approximation.
Both ways give close results. Finding the average spins as functions of time,
we can calculate the radiation intensity.

\section{Radiation intensity}

The intensity of radiation, induced by moving spins, can be calculated in two
ways. One possibility is the classical formula
\be
\label{11}
I(t) = \frac{2\mu_0^2}{3c^3} \left | \sum_j \lgl \ddot{\bS}_j \rgl
\right |^2
\ee
that should provide good approximation for high spins $S \gg 1$. The other way
is to use the quantum formula \cite{Yukalov_5,Yukalov_36,Yukalov_37}, according
to which the radiation intensity
\be
\label{12}
I(t) = I_{inc}(t) + I_{coh}(t)
\ee
is the sum of the incoherent radiation intensity
\be
\label{13}
 I_{inc}(t) = 2\om_0\gm_0 SN [ 1 + s(t) ]
\ee
and the coherent radiation intensity
\be
\label{14}
  I_{coh}(t) = 2\om_0\gm_0 S^2N^2 \vp_0 w(t) \;  ,
\ee
where the natural width is
$$
 \gm_0 \equiv \frac{2}{3} \; | \vec{\mu}|^2 k_0^3  = \frac{1}{3} \; \mu_0^2 k_0^3
\qquad
\left ( k_0 \equiv \frac{\om_0}{c} \right ) \;,
$$
$\vp_0$ is a form-factor, and
\be
\label{15}
 s(t) \equiv \frac{1}{NS} \sum_{j=1}^N \lgl S_j^z(t) \rgl \; , \qquad
 w(t) \equiv \frac{1}{N^2S^2} \sum_{i\neq j}^N \lgl S_i^+(t) S_j^-(t) \rgl \;  .
\ee
If the wavelength is larger than the sample linear size, then $\vp_0 \simeq 1$.
But, when the wavelength is shorter than the system linear size, then the
form-factor essentially depends on the sample shape \cite{Yukalov_37,Allen_38}.

We have accomplished computer simulation for $N$ magnetic nanomolecules
possessing spin $S = 10$, such as Mn$_{12}$ or Fe$_8$, employing the parameters
typical of these nanomolecules, for which $D_2$ and $D_4$ are negligible. The
spin system is prepared in a nonequilibrium initial state, with the external
magnetic field directed along the initial spin polarization, so that the spins
tend to reverse to the opposite direction. It is convenient to consider a
dimensionless radiation intensity, expressed through the units of
$$
I_0 \equiv \frac{2\mu_0^2}{3c^3} \; \gm_2^4 \qquad
\left ( \gm_2 \equiv \frac{1}{T_2} \right ) \;  ,
$$
multiplied by the number of nanomolecules squared, $N^2$, where
$\mu_0 = -2 \mu_B = 1.855 \times 10^{-20}$ erg/G and $\gamma_2 = 10^{10}$ 1/s,
which gives $I_0 = 0.852 \times 10^{-38}$ W. All frequencies are measured in units
of $\gamma_2$. The resonance condition $\omega = \omega_0$ is assumed.

We have considered the influence of different factors on the radiation intensity.
Thus, the role of the Zeeman frequency is exemplified in Fig. 1, showing that the
larger the Zeeman frequency, the higher the radiation intensity. Figure 2
illustrates that the larger the initial spin polarization, the larger the
radiation intensity. Figure 3 shows that increasing the magnetic anisotropy
suppresses the radiation intensity. The role of dipole interactions is described
in Fig. 4, demonstrating that they suppress the radiation intensity by a factor
of $1.5$. In Fig. 5, we study the role of the sample shape and its orientation,
from where it follows that, under the same number of nanomolecules, the most
favorable situation, with the highest radiation intensity, corresponds to the chain
of nanomolecules along the resonator axis.

Calculations for the coherent radiation intensity (14) reduces to the solution
of the evolution equation for the coherence function $w(t)$. Numerical solution
yields the results close to the quasi-classical case, illustrated in Figs. 1 to 5.
This is not surprising, since the coherent regime is known to be well represented
by a quasi-classical approximation. The maximal number of coherently radiating
spins can be estimated as $N_{coh} \sim \rho V_{coh}$, where $\rho$ is the
density of nanomolecules and $V_{coh}$ is the coherence volume. The latter, for
a cylindrical sample, is $V_{coh} \sim \pi R_{coh}^2 L$, where $L$ is the cylinder
length and $R_{coh}$ is a coherence radius \cite{Yukalov_37}, which is of order
$0.3 \sqrt{\lambda L}$. This gives
$$
 N_{coh} \sim \rho\lbd L^2 \;  .
$$
The typical density of magnets, formed by nanomolecules, is
$\rho \approx 0.4 \times 10^{21}$ cm$^{-3}$. For the Zeeman frequency
$\omega \sim 2 \times 10^{13}$ 1/s, the wavelength is $\lambda \sim 10^{-2}$ cm.
If $\lambda \sim L$, then $N_{coh} \sim 10^{14}$. The typical time of a
superradiant pulse is $10^{-11}$ s.

In this way, magnetic nanomolecules and nanoclusters can be described by a
similar macroscopic Hamiltonian. In the process of spin reversal from an
initially prepared non-equilibrium state, there appears spin superradiance,
due to the Purcell effect of the resonator feedback field. The radiation is
mainly absorbed by the resonant coil surrounding the sample.

\section*{Acknowledgments}

The authors are grateful for financial support to the Russian Foundation for
Basic Research (grant 13-02-96018) and to the Perm Ministry of Education
(grant C-26/628).

\vskip 2cm

\begin{figure}[ht]
\hspace{2cm}
\hbox{
\includegraphics[width=6cm]{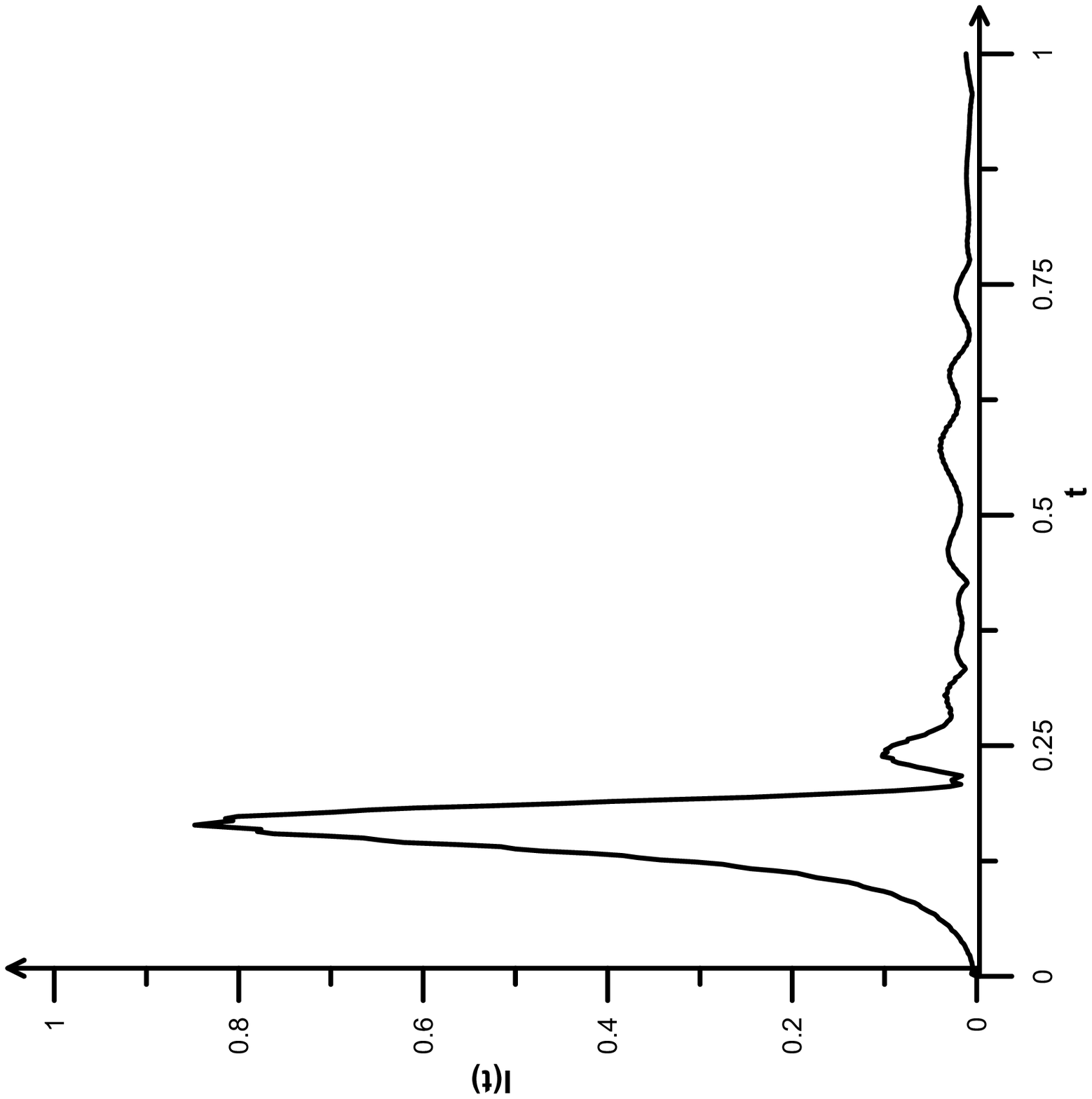} \hspace{1cm}
\includegraphics[width=6cm]{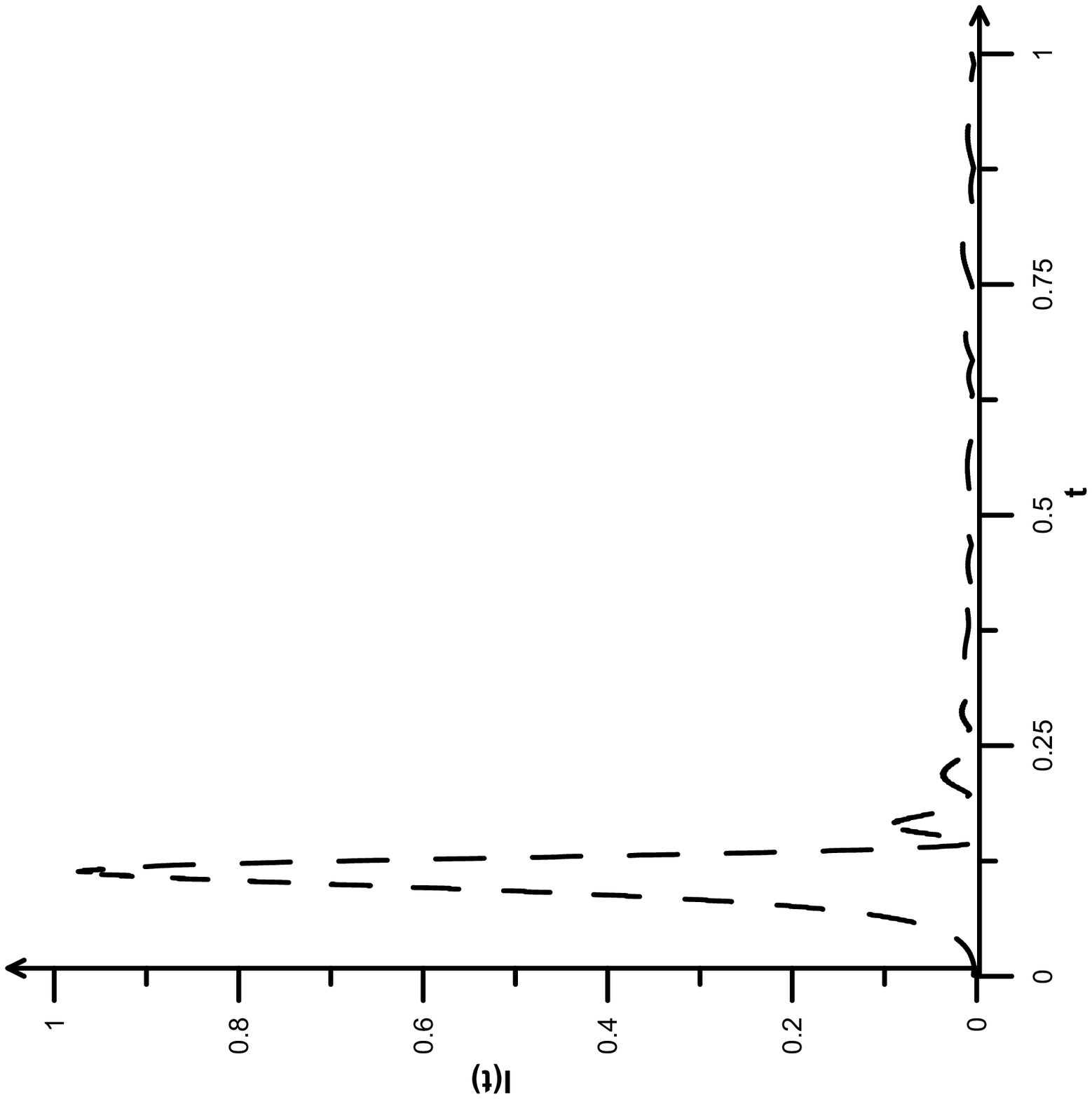}} 
\centerline{ \hspace{-4.7cm}
\hbox{
\includegraphics[width=6cm]{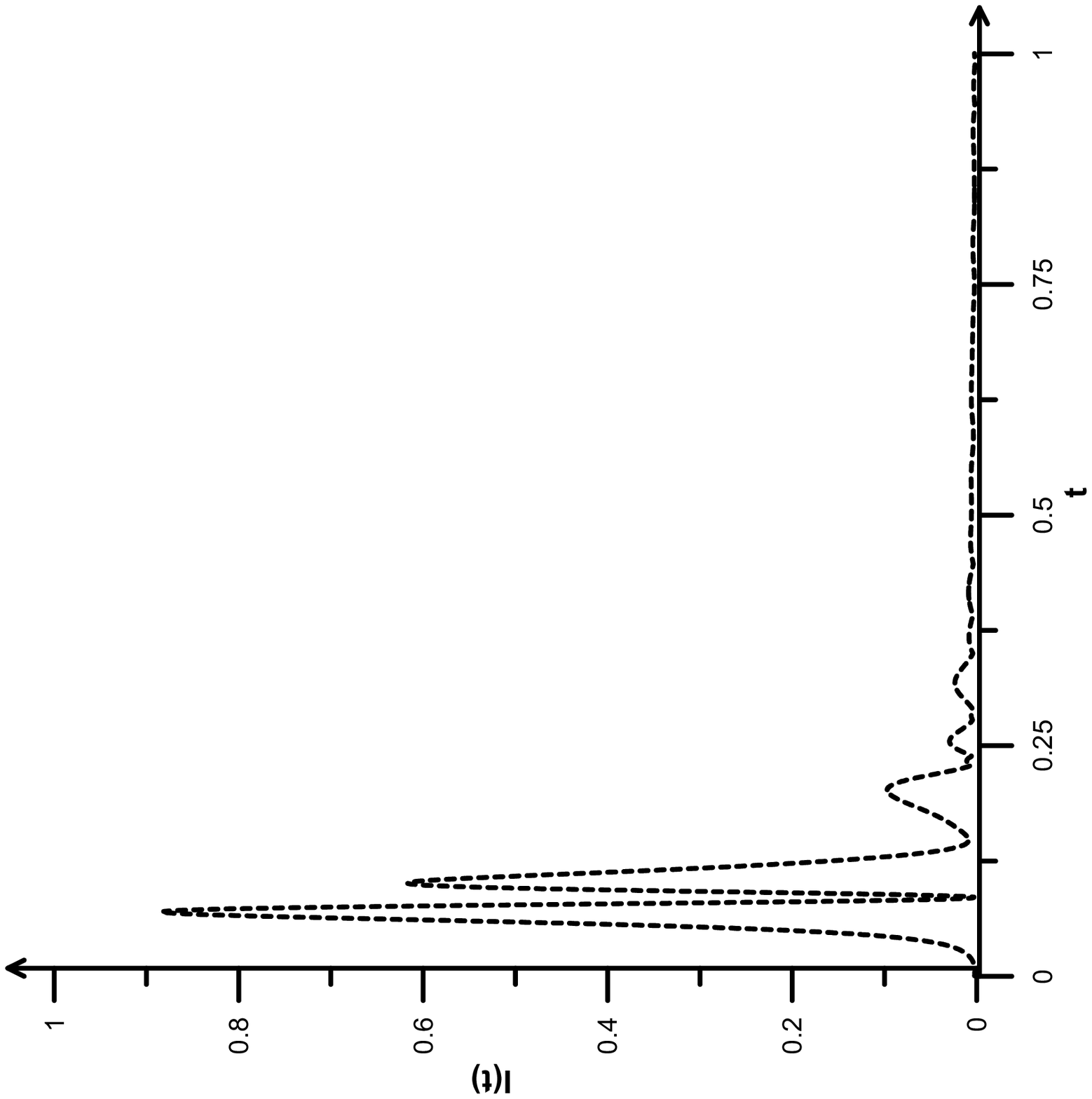}} }
\vspace{1.8cm}
\caption{Radiation intensity (11) from $N = 125$ nanomolecules, with
molecular spin $S = 10$, for a cubic sample. Initial reduced polarization
is $s_0 = 0.9$, the anisotropy frequency is $\omega_D = 20$, and the resonator
damping is $\gamma = 10$. The Zeeman frequency is $\omega_0 = 1000$ (solid line),
with the intensity in units of $3.2 \times 10^{13} N^2 I_0$; $\omega_0 = 2000$
(long-dashed line), in units of $0.8 \times 10^{15} N^2 I_0$, and
$\omega_0 = 5000$ (short-dashed line), in units of $5.1 \times 10^{16} N^2 I_0$. }
\label{fig:Fig.1}
\end{figure}

\begin{figure}[ht]
\hspace{6.5cm}
\includegraphics[width=6.5cm]{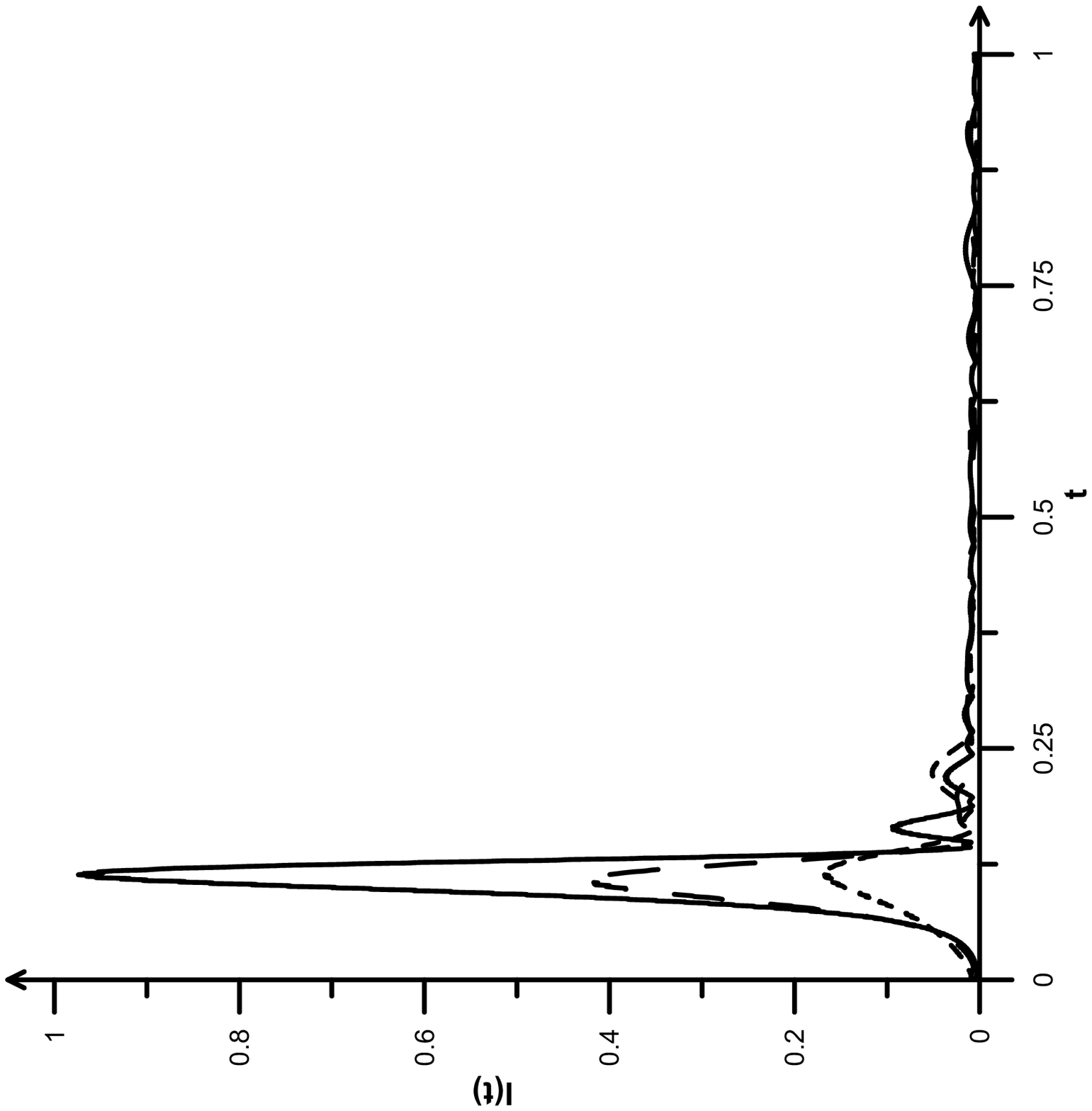}   
\vspace{1.8cm}
\caption{Radiation intensity (11) for a cubic sample of $N =125$ nanomolecules,
with spin $S = 10$, under the Zeeman frequency $\omega_0 = 2000$, anisotropy
frequency $\omega_D = 20$, and the resonator damping $\gamma = 10$, for different
initial polarizations: $s_0 = 0.9$ (solid line), $s_0 = 0.7$ (long-dashed line),
and $s_0 = 0.5$ (short-dashed line). All intensities are in units of
$0.8 \times 10^{15} N^2 I_0$.
}
\label{fig:Fig.2}
\end{figure}

\begin{figure}[ht]
\hspace{6.5cm}
\includegraphics[width=6.5cm]{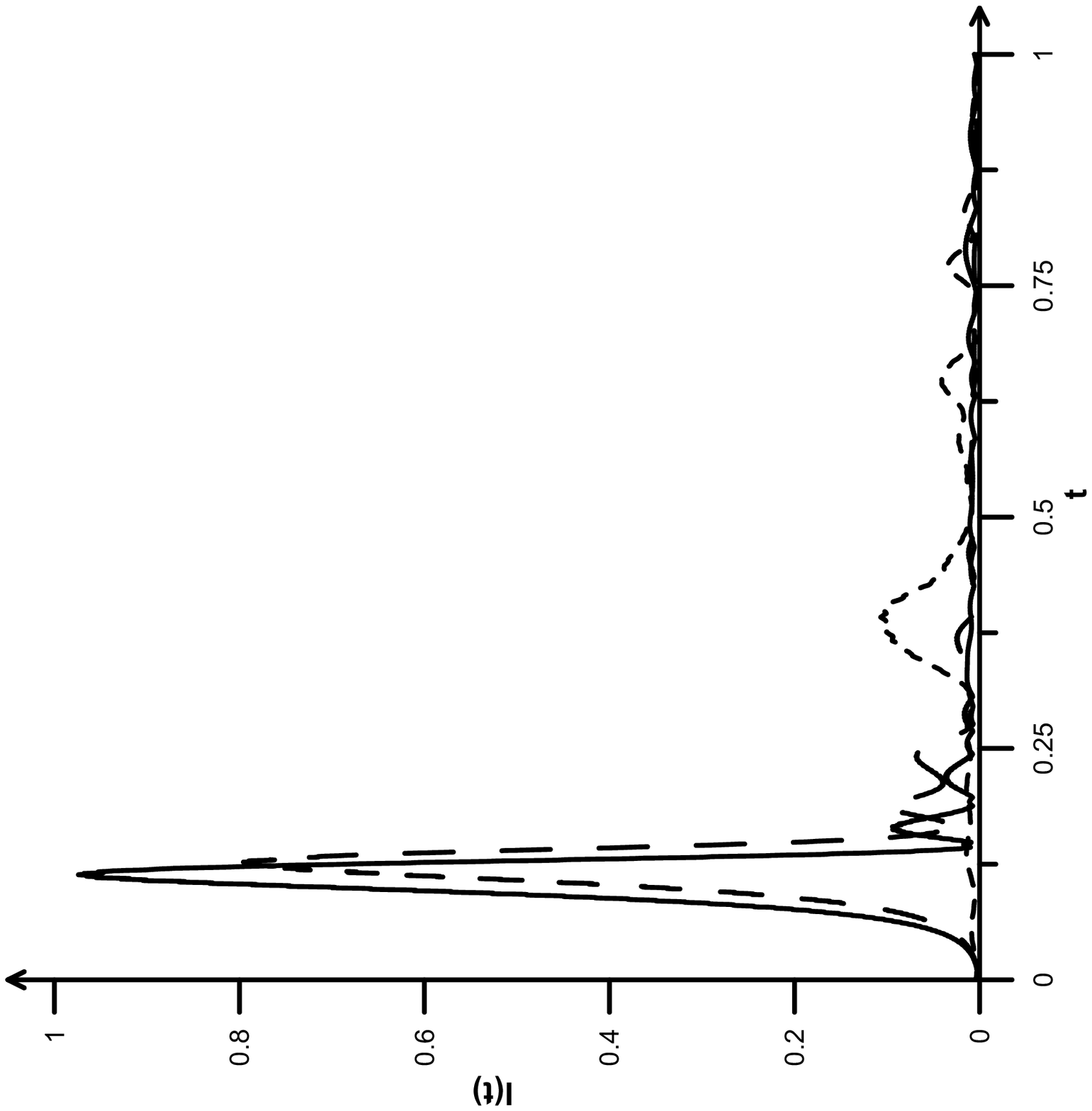}   
\vspace{1.8cm}
\caption{Radiation intensity (11) for a cubic sample of $N = 125$ nanomolecules,
with spin $S = 10$, under the Zeeman frequency $\omega_0 = 2000$, resonator
damping $\gamma = 10$, and the initial spin polarization $s_0 = 0.9$, for varying
anisotropy frequency: $\omega_D = 20$ (solid line), $\omega_D = 50$
(long-dashed line), and $\omega_D = 100$ (short-dashed line). The values of the
radiation intensity are in units of $0.8 \times 10^{15} N^2 I_0$.

}
\label{fig:Fig.3}
\end{figure}

\newpage

\begin{figure}[ht]
\hspace{5cm}
\includegraphics[width=5.93cm]{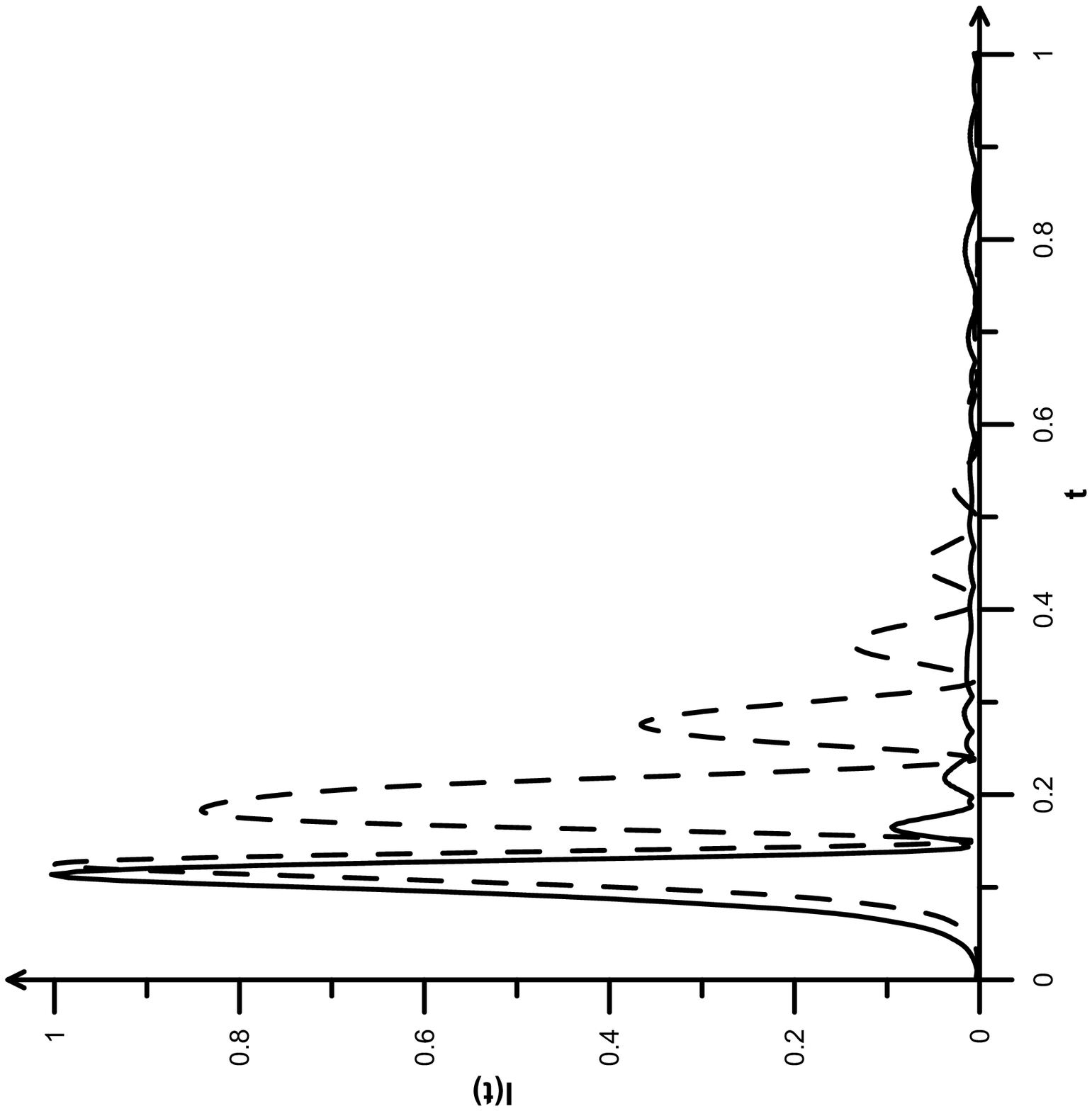} 
\vspace{1.8cm}
\caption{Radiation intensity (11) for a cubic sample of $N = 125$ nanomolecules,
with spin $S = 10$, the Zeeman frequency $\omega_0 = 2000$, anisotropy frequency
$\omega_D = 20$, resonator damping $\gamma = 10$, and the initial spin polarization
$s_0 = 0.9$, for the cases with dipole interactions (solid line), in units of
$0.8 \times 10^{15} N^2 I_0$, and without these interactions (dashed line), in
units of $1.2 \times 10^{15} N^2 I_0$.
}
\label{fig:Fig.4}
\end{figure}

\begin{figure}[ht]
\hspace{6.5cm}
\includegraphics[width=5.93cm]{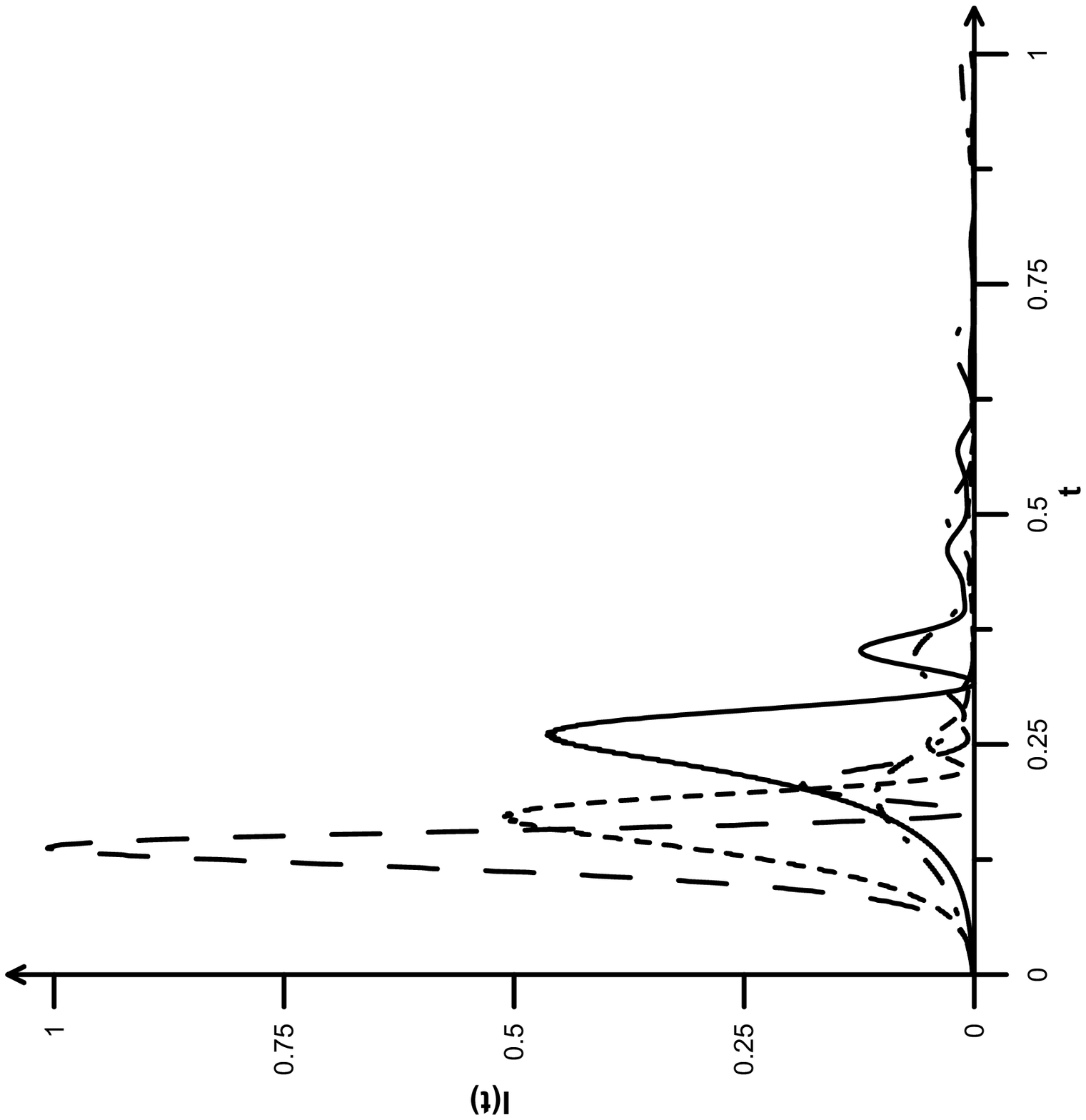} 
\vspace{1.8cm}
\caption{Radiation intensity (11) for a cubic sample of $N = 144$ nanomolecules,
with spin $S = 10$, the Zeeman frequency $\omega_0 = 2000$, anisotropy frequency
$\omega_D = 20$, resonator damping $\gamma = 30$, and the initial spin polarization
$s_0 = 0.9$, for different sample shapes and orientations: the chain of molecules
along the $z$ - axis (solid line), the chain along the $x$ - axis (long-dashed line),
the $y - z$ plane of molecules (short-dashed line), and the $x - y$ plane of
molecules (dotted-dashed line). The intensities are in units of
$1.2 \times 10^{15} N^2 I_0$.
}
\label{fig:Fig.5}
\end{figure}


\newpage


\begin{thebibliography}{99}

\bibitem{Barbara_1}
Barbara B, Thomas L, Lionti F, Chioresku I, and Sulpice A  1999
{\it J. Magn. Magn. Mater.} {\bf 200} 167

\bibitem{Wernsdorfer_2}
Wernsdorfer W  2001
{\it Adv. Chem. Phys.} {\bf 118} 99

\bibitem{Ferre_3}
Ferr\'{e} J  2002
{\it Top. Appl. Phys.} {\bf 83} 127

\bibitem{Yukalov_4}
Yukalov V I  2002
{\it Laser Phys.} {\bf 12} 1089

\bibitem{Yukalov_5}
Yukalov V I and Yukalova E P  2004
{\it Phys. Part. Nucl.} {\bf 35} 348

\bibitem{Bedanta_6}
Bedanta S and Kleemann W  2009
{\it J. Phys. D} {\bf 42} 013001

\bibitem{Berry_7}
Berry C C  2009
{\it J. Phys. D} {\bf 42} 224003

\bibitem{Beveridge_8}
Beveridge J S, Stephens J R, and Williams M E  2011
{\it Annu. Rev. Annal. Chem.} {\bf 4} 251

\bibitem{Hoang_9}
Hoang V V and Canguli D  2012
{\it Phys. Rep.} {\bf 518} 81

\bibitem{Purcell_10}
Purcell E M  1946
{\it Phys. Rev.} {\bf 69} 681

\bibitem{Kiselev_11}
Kiselev J F, Prudkoglyad A F, Shumovsky A S, and Yukalov V I  1988
{\it Mod. Phys. Lett. B} {\bf 1} 409

\bibitem{Belozerova_12}
Belozerova T S, Henner V K, and Yukalov V I  1992
{\it Phys. Rev. B} {\bf 46} 682

\bibitem{Belozerova_13}
Belozerova T S, Henner V K, and Yukalov V I  1992
{\it Laser Phys.} {\bf 2} 545

\bibitem{Yukalov_14}
Yukalov V I  1992
{\it Laser Phys.} {\bf 2} 559

\bibitem{Yukalov_15}
Yukalov V I  1993
{\it Laser Phys.} {\bf 3} 870

\bibitem{Yukalov_16}
Yukalov V I  1995
{\it Phys. Rev. Lett.} {\bf 75} 3000

\bibitem{Yukalov_17}
Yukalov V I  1996
{\it Phys. Rev. B} {\bf 53} 9232

\bibitem{Yukalov_18}
Yukalov V I  1997
{\it Laser Phys.} {\bf 7} 58

\bibitem{Belozerova_19}
Belozerova T S, Davis C L, and Henner V K  1998
{\it Phys. Rev. B} {\bf 58} 3111

\bibitem{Yukalov_20}
Yukalov V I  2005
{\it Phys. Rev. B} {\bf 71} 184432

\bibitem{Davis_39}
Davis C L, Henner V K, Tchernatinsky A V, and Kaganov. I V  2005
{\it Phys. Rev. B} {\bf 72} 054406

\bibitem{Yukalov_40}
Yukalov V I, Henner V K, Kharebov P V, and Yukalova E P  2008
{\it Laser Phys. Lett.} {\bf 5}  887

\bibitem{Yukalov_21}
Yukalov V I, Henner V K, and Kharebov P V  2008
{\it Phys. Rev. B} {\bf 77} 134427

\bibitem{Yukalov_22}
Yukalov V I and Yukalova E P  2011
{\it Laser Phys. Lett.} {\bf 8} 804

\bibitem{Henner_23}
Henner V, Raikher Y, and Kharebov P  2011
{\it Phys. Rev. B} {\bf 84} 144412

\bibitem{Dicke_24}
Dicke R H  1954
{\it Phys. Rev.} {\bf 93} 99

\bibitem{Walther_25}
Walther H, Varcoe B T, Englert B G, and Becker T  2006
{\it Rep. Prog. Phys.} {\bf 69} 1625

\bibitem{Manassah_26}
Manassah J T  2011
{\it Phys. Rev. A} {\bf 83} 025801

\bibitem{Manassah_27}
Manassah J T  2012
{\it Laser Phys.} {\bf 22} 738

\bibitem{Yukalov_28}
Yukalov V I  2005
{\it Laser Phys. Lett.} {\bf 2} 356

\bibitem{Chen_29}
Chen H Y, Lee Y, Bowen S, and Hilty C  2011
{\it J. Magn. Res.} {\bf 208} 204

\bibitem{Krishnan_30}
Krishnan V V and Murali N  2013
{\it Prog. Nucl. Magn. Res. Spectrosc.} {\bf 68} 41

\bibitem{Shavishvili_31}
Shavishvili T M, Khutsishvili K O, Fokina N P, and Lavrentiev G V  1989
{\it J. Tech. Phys. Lett.} {\bf 15} 33

\bibitem{Fokina_32}
Fokina N P and Khutsishvili K O  1990
{\it Phys. Met. Metallogr.} {\bf 69} 65

\bibitem{Nazarova_33}
Nazarova O V, Fokina N P, and Khutsishvili K O  1991
{\it Phys. Met. Metallogr.} {\bf 70} 44

\bibitem{Jin_34}
Jin F, De Raedt H, Yuan S, Katsnelson M I, Miyashita S, and Michielsen K  2010
{\it J. Phys. Soc. Jap.} {\bf 79} 124005

\bibitem{Birman_35}
Birman J L, Nazmitdinov R G, and Yukalov V I  2013
{\it Phys. Rep} {\bf 526} 1

\bibitem{Yukalov_36}
Yukalov V I and Yukalova E P  2005
{\it Laser Phys. Lett.} {\bf 2} 302

\bibitem{Yukalov_37}
Yukalov V I  2014
{\it Laser Phys.} {\bf 24} 094015

\bibitem{Allen_38}
Allen L and Eberly J H  1975
{\it Optical Resonance and Two-Level Atoms} (New York: Wiley)

\end{thebibliography}
\end{document}